\documentclass{naturedraft}

\usepackage{aasymbols}
\usepackage{graphicx}
\usepackage{hyperref}

\hypersetup{
  pdftitle={AR Sco},
  pdfauthor={Tom Marsh, Warwick},
  pdfborder={0 0 0},
  colorlinks=true,
  allcolors=[rgb]{0.0, 0.0, 0.7},
}

\usepackage[style=nature,defernumbers=true,backend=biber]{biblatex}
\defbibheading{blank}{}

\AtEveryBibitem{%
  \clearfield{day}%
  \clearfield{month}%
  \clearfield{endday}%
  \clearfield{endmonth}%
}

\usepackage{siunitx}

\DeclareSIUnit\parsec{pc}
\DeclareSIUnit\mjy{mJy}
\DeclareSIUnit\gauss{G}
\DeclareSIUnit\msun{M_{\ensuremath{\odot}}}
\DeclareSIUnit\rsun{R_{\ensuremath{\odot}}}
\DeclareSIUnit\year{yr}

\newcommand{\lmean}{\bar{L}}
\newcommand{\lstar}{L_{\star}}
\newcommand{\lspin}{L_{\dot{\nu}}}
\newcommand{\lextra}{L_{+}}
\newcommand{\fspin}{\nu_{\mathrm{S}}}
\newcommand{\fbeat}{\nu_{\mathrm{B}}}
\newcommand{\forb}{\nu_{\mathrm{O}}}
\newcommand{\spitzer}{\textit{Spitzer}}
\newcommand{\herschel}{\textit{Herschel}}
\newcommand{\hst}{\textit{HST}}
\newcommand{\wise}{WISE}
\newcommand{\swift}{\textit{Swift}}

\title{A radio pulsing white dwarf binary star}

\author{
T.R. Marsh$^{\ref{a:war}}$,
B.T. G\"ansicke$^{\ref{a:war}}$,
S. H\"{u}mmerich$^{\ref{a:bav},\ref{a:aavso}}$,
F.-J. Hambsch$^{\ref{a:bav},\ref{a:aavso},\ref{a:vvs}}$,
K. Bernhard$^{\ref{a:bav},\ref{a:aavso}}$,
C.Lloyd$^{\ref{a:sux}}$,
E. Breedt$^{\ref{a:war}}$,
E.R. Stanway$^{\ref{a:war}}$,
D.T. Steeghs$^{\ref{a:war}}$,
S.G. Parsons$^{\ref{a:val}}$,
O. Toloza$^{\ref{a:war}}$,
M.R. Schreiber$^{\ref{a:val}}$,
P.G. Jonker$^{\ref{a:sron},\ref{a:nij}}$,
J. van Roestel$^{\ref{a:nij}}$,
T. Kupfer$^{\ref{a:calt}}$,
A.F. Pala$^{\ref{a:war}}$,
V.S. Dhillon$^{\ref{a:she},\ref{a:iac},\ref{a:lla}}$,
L.K. Hardy$^{\ref{a:she}}$,
S.P. Littlefair$^{\ref{a:she}}$,
A. Aungwerojwit$^{\ref{a:nar}}$,
S. Arjyotha$^{\ref{a:crai}}$,
D. Koester$^{\ref{a:kiel}}$,
J.J. Bochinski$^{\ref{a:ou}}$,
C.A. Haswell$^{\ref{a:ou}}$,
P. Frank$^{\ref{a:bav}}$,
P.J. Wheatley$^{\ref{a:war}}$}

\begin{document}

\maketitle

\begin{affiliations}

\item \label{a:war} Department of Physics, Gibbet Hill Road, University of
  Warwick, Coventry, CV4~7AL, UK

\item \label{a:bav} Bundesdeutsche Arbeitsgemeinschaft f\"{u}r
  Ver\"{a}nderliche Sterne e.V. (BAV), Berlin, Germany

\item \label{a:aavso} American Association of Variable Star Observers
  (AAVSO),Cambridge, MA, USA

\item \label{a:vvs} Vereniging Voor Sterrenkunde (VVS), Brugge, Belgium 

\item \label{a:sux} Department of Physics and Astronomy, University of
  Sussex, Brighton, BN1~9QH, UK

\item \label{a:val} Instituto de F\'{i}sica y Astronom\'{i}a, Universidad de
  Valpara\'{i}so, Avenida Gran Bretana 1111, Valpara\'{i}so, Chile 

\item \label{a:sron} SRON, Netherlands Institute for Space Research,
  Sorbonnelaan 2, 3584-CA, Utrecht, The Netherlands

\item \label{a:nij} Department of Astrophysics/IMAPP, Radboud University
  Nijmegen, P.O. box 9010, 6500 GL Nijmegen, The Netherlands

\item \label{a:calt} Division of Physics, Mathematics and Astronomy,
  California Institute of Technology, Pasadena, CA~91125, USA

\item \label{a:she} Department of Physics and Astronomy, University of
  Sheffield, Sheffield, S3~7RH, UK

\item \label{a:iac} Instituto de Astrofisica de Canarias (IAC), E-38205 La
  Laguna, Tenerife, Spain

\item \label{a:lla} Universidad de La Laguna, Dpto.\ Astrofisica, E-38206 La
  Laguna, Tenerife, Spain

\item \label{a:nar} Department of Physics, Faculty of Science, Naresuan
  University, Phitsanulok 65000, Thailand

\item \label{a:crai} Program of Physics, Faculty of Science and Technology,
  Chiang Rai Rajabhat University, Chiang Rai 57100, Thailand

\item \label{a:kiel} Institut f\"{u}r Theoretische Physik und Astrophysik,
  University of Kiel, 24098 Kiel, Germany

\item \label{a:ou} Department of Physical Sciences, The Open University,
  Milton Keynes, UK

\end{affiliations}

\begin{refsegment}
\defbibfilter{notother}{not segment=\therefsegment}

\begin{center}
\href{http://dx.doi.org/10.1038/nature18620}{\textbf{doi:10.1038/nature18620}}
\end{center}

\begin{abstract}
  White dwarfs are compact stars, similar in size to Earth but $\mathbf{\sim
    200}$,$\mathbf{000}$ times more massive$^{1}$.
  Isolated white dwarfs emit most of their power from ultraviolet to
  near-infrared wavelengths, but when in close orbits with less dense stars,
  white dwarfs can strip material from their companions, and the resulting
  mass transfer can generate atomic line$^{2}$ and
  X-ray$^{3}$ emission, as well as near- and
  mid-infrared radiation if the white dwarf is
  magnetic$^{4}$. However, even in binaries, white
  dwarfs are rarely detected at far-infrared or radio frequencies. Here we
  report the discovery of a white dwarf / cool star binary that emits from
  X-ray to radio wavelengths.  The star, AR~Scorpii (henceforth AR~Sco), was
  classified in the early 1970s as a $\delta$-Scuti
  star$^{5}$, a common variety of periodic variable
  star. Our observations reveal instead a $\mathbf{3.56}\,$hr period close
  binary, pulsing in brightness on a period of $\mathbf{1.97}\,$min. The
  pulses are so intense that AR~Sco's optical flux can increase by a factor of
  four within $\mathbf{30}\,$s, and they are detectable at radio frequencies, the
  first such detection for any white dwarf system. They reflect the spin of a
  magnetic white dwarf which we find to be slowing down on a
  $\mathbf{10^7}\,$yr timescale. The spin-down power is an order of magnitude
  larger than that seen in electromagnetic radiation, which, together with an
  absence of obvious signs of accretion, suggests that AR~Sco is primarily
  spin-powered. Although the pulsations are driven by the white dwarf's spin,
  they originate in large part from the cool star.  AR~Sco's broad-band
  spectrum is characteristic of synchrotron radiation, requiring relativistic
  electrons. These must either originate from near the white dwarf or be
  generated \textit{in situ} at the M star through direct interaction with the
  white dwarf's magnetosphere.
\end{abstract}

\begin{figure}
  \begin{center}
    \includegraphics[width=0.75\textwidth]{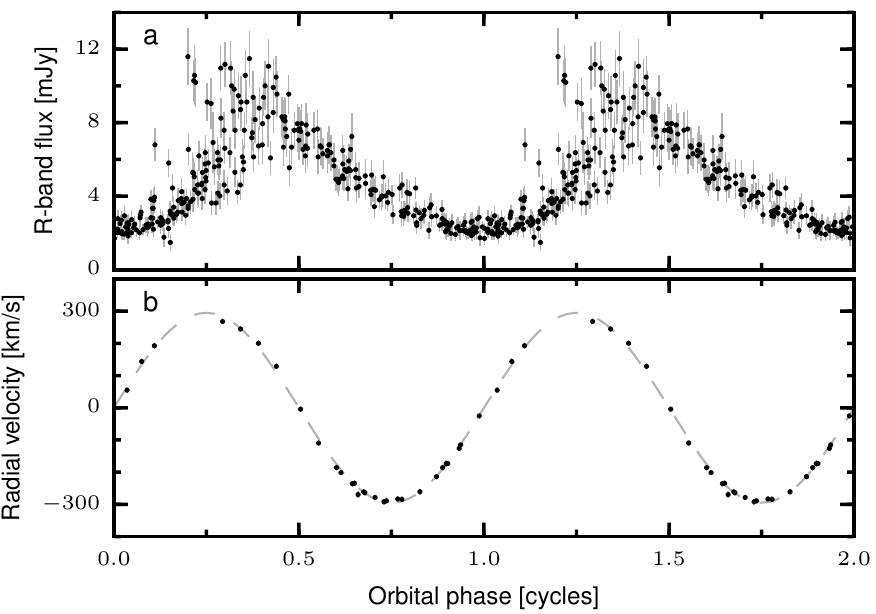}
  \end{center}
  \caption{\textbf{AR~Sco's optical brightness and radial velocity curve.}
    \textbf{a}, Photometry ($\SI{30}{\s}$ exposures) taken over 7 years shows
    a factor four variation in brightness on a $\SI{3.56}{\hour}$ period, with
    large scatter at some phases. \textbf{b}, The M star varies sinusoidally
    in velocity on the same period, showing it to be the orbital period of a
    close, circular orbit binary star. The orbital phase is defined so that at
    phase~0 the M star is at its closest point to Earth. $\pm 1\sigma$ error
    bars are shown, but are too small to see in \textbf{b}.\label{f:rvs}}
\end{figure}

\begin{figure}
\begin{center}
\includegraphics[width=0.99\textwidth]{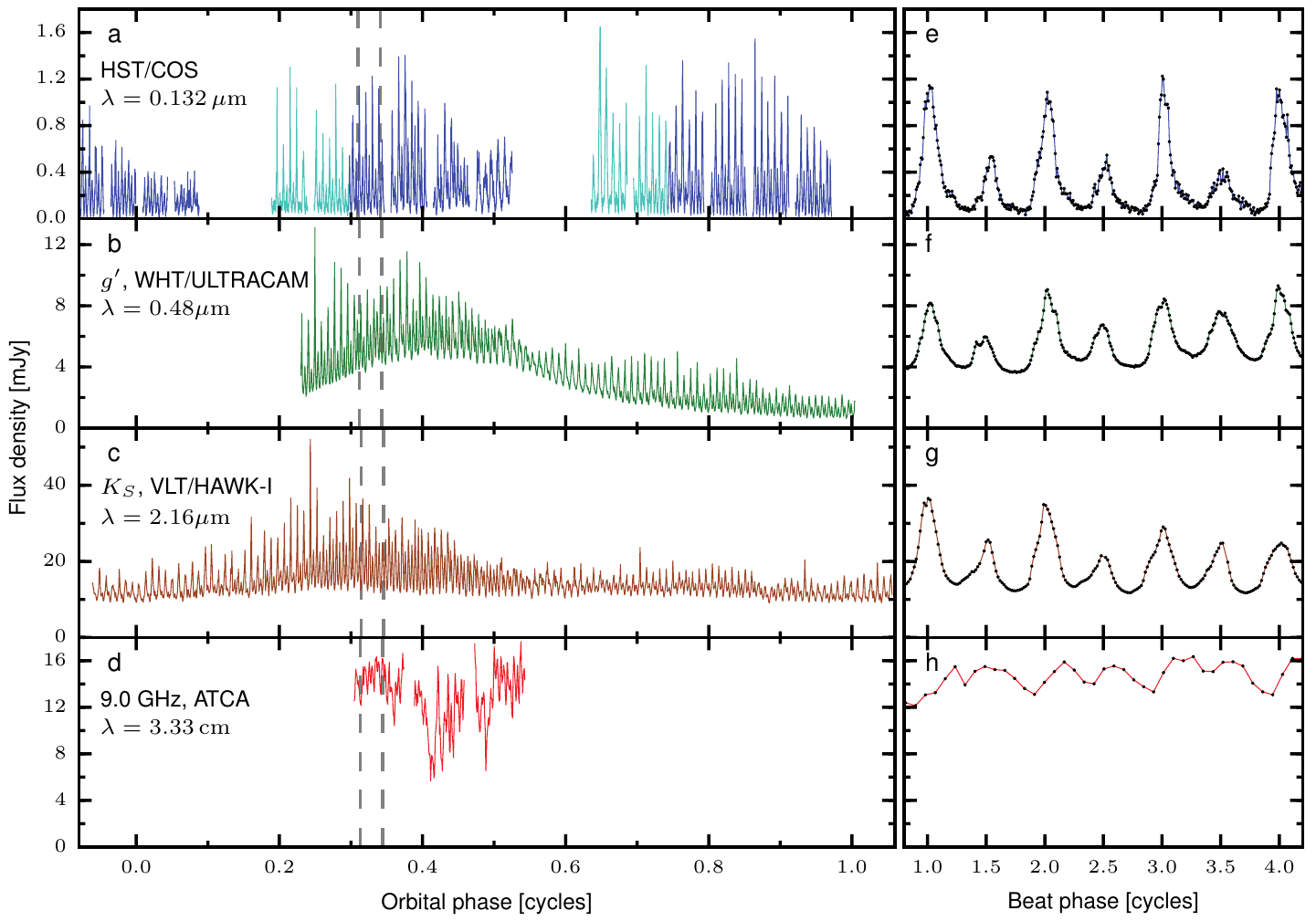}
\end{center}
\caption{\textbf{Ultraviolet, optical, infrared and radio fluxes of AR Sco.}
  \textbf{a}--\textbf{d}, High-speed measurements of the UV, optical, infrared
  and radio fluxes of AR~Sco plotted against orbital phase. Sections of
  similar orbital phases, marked by dashed lines, are shown expanded in
  \textbf{e}--\textbf{h} where they are plotted against the beat pulsation
  phase. Black dots mark individual measurements. None of the four sets of
  data were taken simultaneously in time. The different colours in \textbf{a}
  indicate that the data were acquired in different orbital
  cycles.\label{f:lc}}
\end{figure}

\begin{figure}
\begin{center}
\includegraphics[width=0.99\textwidth]{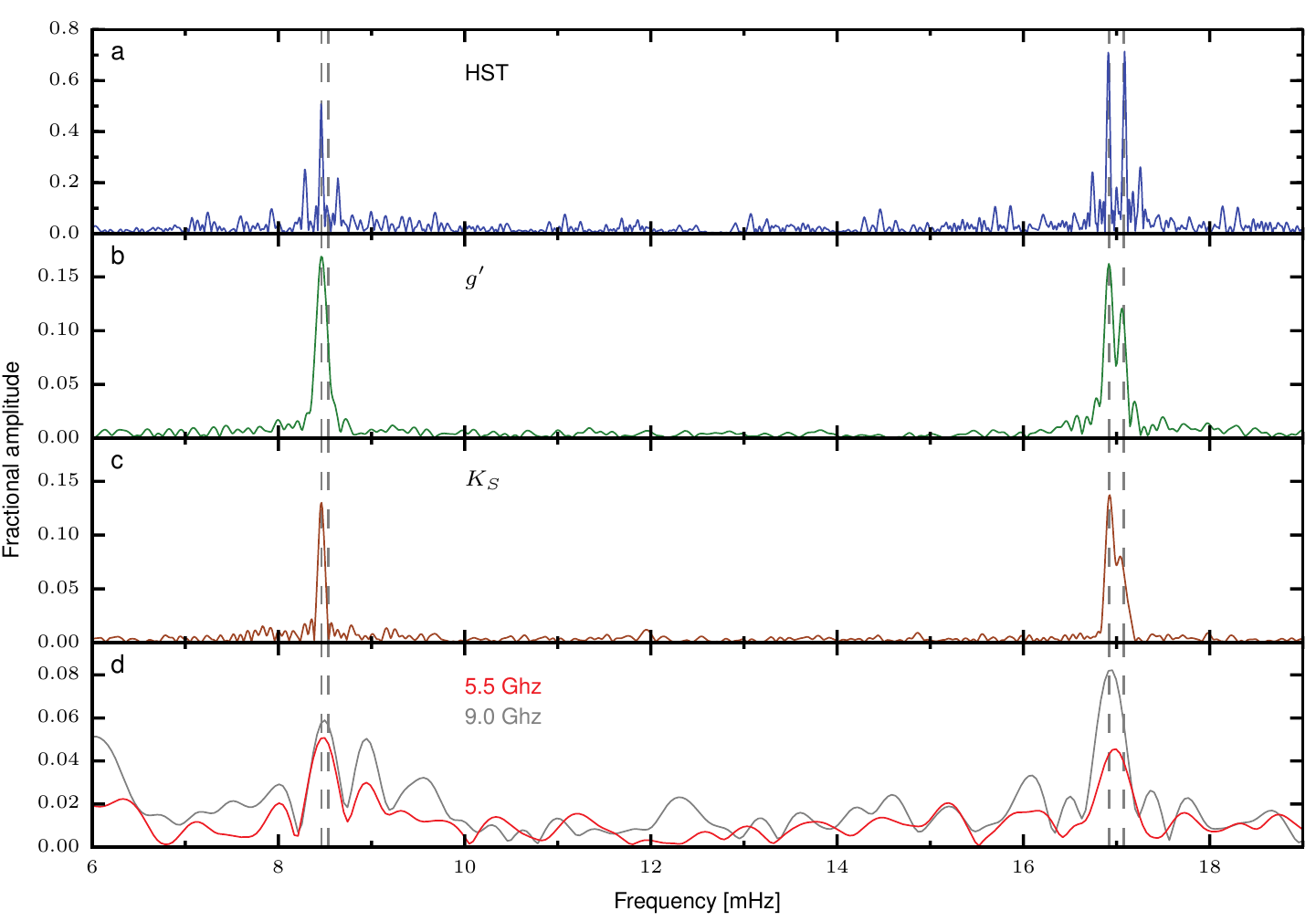}
\end{center}
\caption{\textbf{Fourier amplitudes of the ultraviolet, optical, infrared and
    radio fluxes of AR~Sco versus temporal frequency .}
  \textbf{a}--\textbf{d} are the amplitude spectra corresponding to
  \textbf{a}--\textbf{d} of the light-curves of Fig.~\protect\ref{f:lc}.  All
  bands show signals with a fundamental period of $\sim \SI{1.97}{\minute}$
  ($\SI{8.46}{\mHz}$) and its second harmonic.  The signals have two
  components, clearest in the harmonic, which we identify as the spin
  frequency $\fspin$ and ``beat'' frequency $\fbeat = \fspin - \forb$, where
  $\forb$ is the orbital frequency. The beat component is the stronger of the
  two and defines the dominant $\SI{1.97}{\minute}$ pulsation period; the spin
  period is $\SI{1.95}{\minute}$.\label{f:pgram}}
\end{figure}

AR~Sco's brightness varies on a $\SI{3.56}{\hour}$ period
(Fig.~\ref{f:rvs}a); it was this that caused the $\delta$-Scuti
classification$^{5}$. The scatter visible in
Fig.~\ref{f:rvs}a prompted us to take optical photometry with the high-speed
camera ULTRACAM$^{6}$. These data and follow-up
observations taken in the ultraviolet and near-infrared (Extended Data
Table~\ref{t:obslog}) all show strong double-humped pulsations on a
fundamental period of $\SI{1.97}{\minute}$ (Figs~\ref{f:lc} and
\ref{f:pgram}); the scatter in Fig.\ref{f:rvs}a is the result of the
pulsations. Most unusually of all, an hour-long observation at radio
frequencies with the Australia Telescope Compact Array (ATCA) also shows the
pulsations (Figs~\ref{f:lc}d, \ref{f:lc}e and \ref{f:pgram}d). The pulse
fraction,
$(f_{\mathrm{max}}-f_{\mathrm{min}})/(f_{\mathrm{max}}+f_{\mathrm{min}})$,
exceeds $\SI{95}{\percent}$ in the far ultraviolet (Fig.~\ref{f:lc}), and
is still $\SI{10}{\percent}$ at $\SI{9}{\GHz}$ in the radio. Only in X-rays
did we not detect pulses (pulse fraction $<\SI{30}{\percent}$ at
$\SI{99.7}{\percent}$ confidence). AR~Sco's optical magnitude ($g'$) varies
from $16.9$ at its faintest to $13.6$ at its peak, a factor of 20 in flux.

We acquired optical spectra which show a cool M-type main-sequence star
(Extended Data Fig.~\ref{f:spectype}) with absorption lines that change radial
velocity sinusoidally on the $\SI{3.56}{\hour}$ period with amplitude
$K_2=\SI{295\pm4}{\km\per\s}$ (Fig.~\ref{f:rvs}b; we use subscripts ``1'' and
``2'' to indicate the compact star and the M star respectively). The
$\SI{3.56}{\hour}$ period is therefore the orbital period of a close binary
star. The M star's radial velocity amplitude sets a lower limit on the mass of
its companion of $M_1\ge\SI{0.395\pm0.016}{\msun}$. The compact object is not
visible in the spectra, consistent with either a white dwarf or a neutron
star, the only two types of object which can both support a misaligned
magnetic dipole and spin fast enough to match the pulsations.  The optical and
ultraviolet spectra show atomic emission lines (Extended Data
Figs.~\ref{f:spectype} and \ref{f:hstspec}) which originate from the
side of the M star facing the compact object (Extended Data
Fig.~\ref{f:trail}). Their velocity amplitude relative to the M star sets a
lower limit upon the mass ratio $q=M_2/M_1>0.35$ (Extended Data
Fig.~\ref{f:roche}).  This, along with the requirement that the M star fits
within its Roche lobe, defines mass ranges for each star of
$\SI{0.81}{\msun}<M_1\la\SI{1.29}{\msun}$ and
$\SI{0.28}{\msun}<M_2\la\SI{0.45}{\msun}$. The M star's spectral type (M5)
suggests that its mass lies at the lower end of the allowed range for $M_2$.
Assuming that the M star is close to its Roche lobe, its brightness leads to a
distance estimate $d=(M_2/\SI{0.3}{\msun})^{1/3} \SI{116\pm16}{\parsec}$.

The amplitude spectra of the pulsations show the presence of two components of
similar frequency (Fig.~\ref{f:pgram}). Using our own monitoring and
archival optical data spanning 7 years$^{7}$, we
measured precise values for the frequencies of these components, finding their
difference to be within 20 parts per million of the orbital
frequency, $\forb$ (Extended Data Figs~\ref{f:ClearAmps} and \ref{f:CRTSamps},
Extended Data Table~\ref{t:freqs}). The natural interpretation is that the
higher frequency component represents the spin frequency $\fspin$ of the
compact star ($P_{\mathrm{S}}=\SI{1.95}{\minute}$), while its lower frequency
and generally stronger counterpart is a re-processed or ``beat'' frequency
$\fbeat = \fspin - \forb$ ($P_{\mathrm{B}}=\SI{1.97}{\minute}$), assuming that
the compact star spins in the same sense as the binary orbit.

\begin{figure}
\begin{center}
\includegraphics[width=0.99\textwidth]{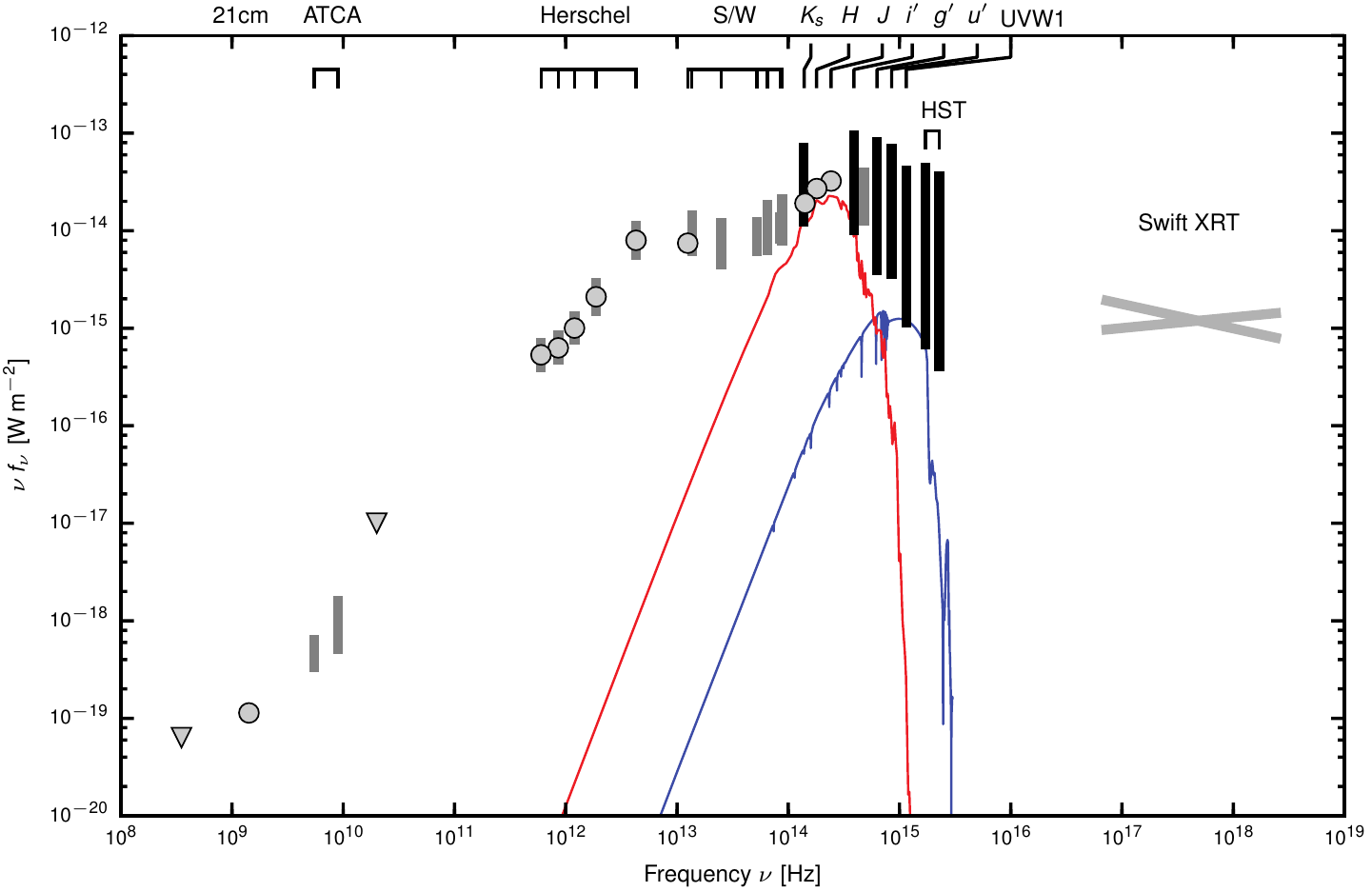}
\end{center}
\caption{\textbf{The wide band Spectral Energy Distribution (SED) of AR Sco.}
  Black bars show the range spanned by intensive, time-resolved data; grey
  bars represent more limited datasets spanning less than the full
  variation. Grey points with error bars ($1\sigma$) represent single
  exposures. The grey lines represent the $\pm 1\sigma$ range of X-ray
  spectral slopes. Triangles are upper-limits. ``S/W'' marks data from
  \spitzer\ and \wise. The red and blue lines show model atmospheres,
  extended at long wavelengths with black-body spectra, for the M star ($R_2 =
  \SI{0.36}{\rsun}$, $T_2 = \SI{3100}{\K}$) and white dwarf ($R_1 =
  \SI{0.01}{\rsun}$, $T_2 = \SI{9750}{\K}$ upper limit) at a distance $d =
  \SI{116}{\parsec}$. See Extended Data Tables~\ref{t:obslog} and
  \ref{t:archive} for details of data sources.\label{f:sed}}
\end{figure}

AR~Sco emits across the electromagnetic spectrum (Fig.~\ref{f:sed}, Extended
Data Table~\ref{t:archive}), and, in the infrared and radio in particular, is
orders of magnitudes brighter than the thermal emission from its component
stars represented by model
atmospheres$^{8,9}$ in
Fig.~\ref{f:sed}. Integrating over the spectral energy distribution (SED)
shown in Fig.~\ref{f:sed} and adopting a distance of $\SI{116}{\parsec}$, we
find a maximum luminosity of $\approx\SI{6.3e25}{\W}$ and a mean of
$\lmean\approx\SI{1.7e25}{\W}$, well in excess of the combined luminosities of
the stellar components $\lstar=\SI{4.4e24}{\W}$. The two possible sources of
this luminosity are accretion and spin-down power of the compact object. A
spinning object of moment of inertia $I$ loses energy at a rate
$\lspin=-4\pi^2I\fspin\dot{\fspin}$ where $\fspin$ and $\dot{\fspin}$ are the
spin frequency and its time derivative. Using the archival optical data we
measured the spin frequency to be slowing, with a frequency derivative of
$\dot{\fspin}=-\SI{2.86(36)e-17}{\hertz\per\s}$.  For parameters typical of
neutron stars and white dwarfs ($M_{\mathrm{NS}}=\SI{1.4}{\msun}$,
$R_{\mathrm{NS}}=\SI{10}{\km}$; $M_{\mathrm{WD}}=\SI{0.8}{\msun}$,
$R_{\mathrm{WD}}=\SI{0.01}{\rsun}$), this leads to
$\lspin(\mathrm{NS})=\SI{1.1e21}{\W}$ and
$\lspin(\mathrm{WD})=\SI{1.5e26}{\W}$. Compared to the mean luminosity in
excess of the stellar contributions, $\lextra=\lmean-\lstar=\SI{1.3e25}{\W}$,
this shows that spin-down luminosity is sufficient to power the system if the
compact object is a white dwarf but not if it is a neutron star. Accretion is
the only possible power source in the case of a neutron star -- an accretion
rate of $\dot{M}_{\mathrm{NS}}=\SI{1.0e-14}{\msun\per\year}$
suffices. Accretion could partially power a white dwarf, but it cannot be the
main source because the rate required,
$\dot{M}_{\mathrm{WD}}=\SI{1.3e-11}{\msun\per\year}$, is high enough that we
should see Doppler-broadened emission lines from the accreting gas whereas
AR~Sco only shows features from the M star.

The observations point toward a white dwarf as the compact object. First,
AR~Sco's distance of $\SI{116}{\parsec}$ is an order of magnitude closer than
the nearest accreting neutron star known,
Cen~X-4$^{10}$, but typical of white dwarf /
main-sequence binaries (closer systems are
known$^{11}$). Second, AR~Sco's X-ray luminosity,
$L_X=\SI{4.9e23}{\W}$, is only $\SI{4}{\percent}$ of the largely-optical
luminosity excess, $\lextra$. By contrast, the X-ray luminosities of accreting
neutron stars are typically 100 times their optical
luminosities$^{12}$. Third, at $P_{\mathrm{S}} =
\SI{1.95}{\minute}$, AR~Sco has a spin period an order of magnitude longer
than any (neutron star powered) radio pulsar
known$^{13}$. Finally, the upper limit masses $M_1 =
\SI{1.29}{\msun}$ and $M_2=\SI{0.45}{\msun}$ are simultaneously low for a
neutron star but high for an M5 M star. A $\SI{0.8}{\msun}$ white dwarf with a
$\SI{0.3}{\msun}$ M dwarf is a more natural pairing.

AR~Sco's observational properties are unique. It may represent an evolutionary
stage of a class of stars known as intermediate polars (IPs), which feature
spinning magnetic white dwarfs accreting from low-mass stars in close
binaries$^{14}$. Only one IP, AE~Aquarii (AE~Aqr), has
a broad-band SED similar to AR~Sco$^{15}$ and
comparably strong radio emission$^{16}$, although it
shows no radio pulsations$^{17}$
($<\SI{0.8}{\percent}$) and its $\SI{0.4}{\percent}$ optical pulsations
compare with $\SI{70}{\percent}$ in AR~Sco. With a $\SI{25}{\percent}$ pulse
fraction, even the IP with the strongest-known optical pulsations,
FO~Aquarii$^{18}$, falls well short of AR~Sco. A key
difference is perhaps the lack of significant accretion in AR~Sco compared to
the IPs. This can be seen from its X-ray luminosity which is less than
$\SI{1}{\percent}$ of the X-ray luminosity of a typical
IP$^{19}$, but above all from its optical and
ultraviolet emission lines which come entirely from the irradiated face of the
M star.  IPs by contrast show Doppler-broadened line emission, often from
accretion discs, and even AE~Aqr, which is in an unusual ``propeller'' state
in which transferred matter is expelled upon encountering the magnetosphere of
its rapidly-spinning $P_{\mathrm{S}} = \SI{33}{\s}$ white
dwarf$^{20,21}$,
shows broad and stochastically variable line emission. We can find no analogue
of AR~Sco's radio properties. Pulsed radio emission has been detected from
brown dwarfs and M stars$^{22,23}$,
but the broad-band nature of AR~Sco's emission, its short pulsation period,
and lack of circular polarisation (our ATCA data constrain it to $<
\SI{10}{\percent}$), distinguish it from these sources.

The white dwarf in AR~Sco is currently spinning down on a timescale $\tau =
\nu/\dot{\nu} = \SI{e7}{\year}$. White dwarfs are not born spinning
rapidly$^{24}$, and a prior stage of accretion-driven
spin-up is required. Depending upon the distance at which the accreting
material coupled to the white dwarf's magnetic field, between
$\SI{0.002}{\msun}$ and $\SI{0.015}{\msun}$ of matter are required to reach
$P_{\mathrm{S}}=\SI{1.95}{\minute}$. For an accretion rate of
$\SI{e-9}{\msun\per\year}$, typical of similar period systems, this takes from
$\SI{2e7}{\year}$ to $\SI{1.5e8}{\year}$. Both spin-up and spin-down
timescales are much shorter than the likely age of the system: the cooling age
of the white dwarf alone exceeds
$\SI{1.2e9}{\year}$$^{25}$. Thus we could be seeing
one of many such episodes in AR~Sco's history. There is empirical evidence for
similar cycling of accretion rate in both white
dwarf$^{26,27}$ and neutron star
binary systems$^{28,29}$.  If so,
since the spin-up and spin-down timescales are similar in magnitude, there
would be a good chance of catching the spin-down phase.

AR~Sco's extremely broad-band SED is indicative of synchrotron emission from
relativistic electrons. A significant fraction appears to come from the cool M
star. We infer this from the dominant beat frequency component that in the
absence of accretion can only come from the M star. Since the M star occupies
$\sim 1/40^{\mathrm{th}}$ of the sky as seen from the white dwarf, while
the spin-down luminosity is $\sim 11.5$ times the mean electromagnetic power,
this requires a mechanism to transfer energy from the white dwarf to the M
dwarf which is more than $40/11.5 = 3.5$ times more efficient than
the interception of isotropically-emitted radiation. At the same time, direct
pulsed emission from the white dwarf must not overwhelm the re-processed
component. Two possibilities are collimated fast particle outflows and direct
interaction of the white dwarf's magnetosphere with the M dwarf, but the exact
emission mechanism operative in AR~Sco is perhaps its most mysterious feature.

\newpage
\noindent
\textbf{References}
\begin{enumerate}
\item Althaus, L.~G., Althaus, L.~G., Isern, J. \& Garc\'ia-Berro, E. Evolutionary and pulsational properties of white dwarf stars. \textit{Astron.\ \& Astrophys.\ Review} \textbf{18,} 471--566 (2010).

\item Szkody, P. \textit{et al.} Cataclysmic Variables from the Sloan Digital Sky Survey. VIII. The Final Year (2007-2008). \textit{Astron.\ J.\ } \textbf{142,} 181--189 (2011).

\item Revnivtsev, M., Revnivtsev, M., Revnivtsev, M., Ritter, H. \& Sunyaev, R. Properties of the Galactic population of cataclysmic variables in hard X-rays. \textit{Astron.\ \& Astrophys.\ } \textbf{489,} 1121--1127 (2008).

\item Parsons, S.~G. \textit{et al.} A magnetic white dwarf in a detached eclipsing binary. \textit{Mon.\ Not.\ R.\ Astron.\ Soc.\ } \textbf{436,} 241--252 (2013).

\item Satyvaldiev, V. On seventeen variable stars. \textit{Astronomicheskij Tsirkulyar} \textbf{633,} 7--8 (1971).

\item Dhillon, V.~S. \textit{et al.} ULTRACAM: an ultrafast, triple-beam CCD camera for high-speed astrophysics. \textit{Mon.\ Not.\ R.\ Astron.\ Soc.\ } \textbf{378,} 825--840 (2007).

\item Drake, A.~J. \textit{et al.} First Results from the Catalina Real-Time Transient Survey. \textit{Astrophys.\ J.\ } \textbf{696,} 870--884 (2009).

\item Koester, D. White dwarf spectra and atmosphere models . \textit{Mem.\ della Soc.\ Astron.\ Italiana} \textbf{81,} 921--931 (2010).

\item Husser, T.-O. \textit{et al.} A new extensive library of PHOENIX stellar atmospheres and synthetic spectra. \textit{Astron.\ \& Astrophys.\ } \textbf{553,} A6: 1--9 (2013).

\item Chevalier, C., Chevalier, C., Chevalier, C., Pedersen, H. \& van der Klis, M. Optical studies of transient low-mass X-ray binaries in quiescence. I - Centaurus X-4: Orbital period, light curve, spectrum and models for the system. \textit{Astron.\ \& Astrophys.\ } \textbf{210,} 114--126 (1989).

\item Thorstensen, J.~R., L\'epine, S. \& Shara, M. Parallax and Distance Estimates for Twelve Cataclysmic Variable Stars. \textit{Astron.\ J.\ } \textbf{136,} 2107--2114 (2008).

\item Bradt, H.~V.~D. \& McClintock, J.~E. The optical counterparts of compact galactic X-ray sources. \textit{Ann.\ Rev.\ Astron.\ \& Astrophys.\ } \textbf{21,} 13--66 (1983).

\item Manchester, R.~N., Manchester, R.~N., Teoh, A. \& Hobbs, M. The Australia Telescope National Facility Pulsar Catalogue. \textit{Astron.\ J.\ } \textbf{129,} 1993--2006 (2005).

\item Patterson, J. The DQ Herculis stars. \textit{Publ.\ Astron.\ Soc.\ Pacif.\ } \textbf{106,} 209--238 (1994).

\item Oruru, B. \& Meintjes, P.~J. X-ray characteristics and the spectral energy distribution of AE Aquarii. \textit{Mon.\ Not.\ R.\ Astron.\ Soc.\ } \textbf{421,} 1557--1568 (2012).

\item Bookbinder, J.~A. \& Lamb, D.~Q. Discovery of radio emission from AE Aquarii. \textit{Astrophys.\ J.\ Lett.\ } \textbf{323,} L131--L135 (1987).

\item Bastian, T.~S., Beasley, A.~J. \& Bookbinder, J.~A. A Search for Radio Pulsations from AE Aquarii. \textit{Astrophys.\ J.\ } \textbf{461,} 1016--1020 (1996).

\item Patterson, J. \& Steiner, J.~E. H2215-086 -King of the DQ Herculis stars. \textit{Astrophys.\ J.\ Lett.\ } \textbf{264,} L61--L64 (1983).

\item Pretorius, M.~L. \& Mukai, K. Constraints on the space density of intermediate polars from the Swift-BAT survey. \textit{Mon.\ Not.\ R.\ Astron.\ Soc.\ } \textbf{442,} 2580--2585 (2014).

\item Wynn, G.~A., King, A.~R. \& Horne, K. A magnetic propeller in the cataclysmic variable AE Aquarii. \textit{Mon.\ Not.\ R.\ Astron.\ Soc.\ } \textbf{286,} 436--446 (1997).

\item Meintjes, P.~J. \& Venter, L.~A. The diamagnetic blob propeller in AE Aquarii and non-thermal radio to mid-infrared emission. \textit{Mon.\ Not.\ R.\ Astron.\ Soc.\ } \textbf{360,} 573--582 (2005).

\item Berger, E. \textit{et al.} Discovery of radio emission from the brown dwarf LP944-20. \textit{Nature} \textbf{410,} 338--340 (2001).

\item Hallinan, G. \textit{et al.} Periodic Bursts of Coherent Radio Emission from an Ultracool Dwarf. \textit{Astrophys.\ J.\ Lett.\ } \textbf{663,} L25--L28 (2007).

\item Charpinet, S., Fontaine, G. \& Brassard, P. Seismic evidence for the loss of stellar angular momentum before the white-dwarf stage. \textit{Nature} \textbf{461,} 501--503 (2009).

\item Renedo, I. \textit{et al.} New Cooling Sequences for Old White Dwarfs. \textit{Astrophys.\ J.\ } \textbf{717,} 183--195 (2010).

\item Hessman, F.~V., G\"ansicke, B.~T. \& Mattei, J.~A. The history and source of mass-transfer variations in AM Herculis. \textit{Astron.\ \& Astrophys.\ } \textbf{361,} 952--958 (2000).

\item Manser, C.~J. \& G\"ansicke, B.~T. Spectroscopy of the enigmatic short-period cataclysmic variable IR Com in an extended low state. \textit{Mon.\ Not.\ R.\ Astron.\ Soc.\ } \textbf{442,} L23--L27 (2014).

\item Archibald, A.~M. \textit{et al.} A Radio Pulsar/X-ray Binary Link. \textit{Science} \textbf{324,} 1411--1414 (2009).

\item Papitto, A. \textit{et al.} Swings between rotation and accretion power in a binary millisecond pulsar. \textit{Nature} \textbf{501,} 517--520 (2013).

\end{enumerate}


\begin{addendum}
\item[Acknowledgements] TRM, ERS, DS, EB, PJW, VSD, SPL and ULTRACAM were
  supported by the Science and Technology Facilities Council (STFC, ST/L000733
  and ST/M001350/1). BTG, AP and PGJ acknowledge support from the European
  Research Council (ERC, 320964 and 647208). OT, SGP and MRS acknowledge
  support from Fondecyt (3140585 and 1141269). MRS also received support from
  Millenium Nucleus RC130007 (Chilean Ministry of Economy). AA acknowledges
  support from the Thailand Research Fund (MRG5680152) and the National
  Research Council of Thailand (R2559B034). Based on observations collected
  with telescopes of the Isaac Newton Group in the Spanish Observatorio del
  Roque de los Muchachos of the Instituto de Astrof\'{i}sica de Canarias, the
  European Organisation for Astronomical Research in the Southern Hemisphere
  (095.D-0489, 095.D-0739, 095.D-0802), the NASA/ESA Hubble Space Telescope
  (14470), and the Thai National Telescope. Archival data from the \herschel,
  \spitzer\ and \wise\ space observatories, and from the Catalina Sky Survey
  were used. We thank the \swift\ mission PI for a target-of-opportunity
  program on AR~Sco with the XRT and UVOT instruments, and the Director of
  ATCA for the award of Director's Discretionary Time.

\item[Competing Interests] The authors declare that they have no
competing financial interests.

\item[Author contributions] TRM organised observations, analysed the data, 
  interpreted the results and was the primary author
  of the manuscript. BTG, AFP, EB, SGP, PGJ, JvR, TK, MRS, and OT acquired,
  reduced and analysed optical and ultraviolet spectroscopy.  ERS
  acquired, reduced and analysed the ATCA radio data. SH, FJH, KB, CL and PF
  first identified the unusual nature of AR~Sco and started the optical
  monitoring campaign. VSD, LKH, SPL, AA, SA, JJB and CAH acquired and reduced the
  high-speed optical photometry. DTS and PJW acquired and analysed \swift\ and
  archival X-ray data. DK calculated the white dwarf model atmosphere. All
  authors commented on the manuscript.

 \item[Correspondence] Correspondence and requests for materials
should be addressed to TRM. \\ (email: t.r.marsh@warwick.ac.uk).
\end{addendum}


\newrefsegment

\newpage

\begin{methods}

\subsection{Data sources.}
AR~Sco's location in the ecliptic plane, not far from the Galactic centre and
only $2.5^\circ$ North-West of the centre of the Ophiuchus molecular cloud,
means that it appears in many archival observations. It is detected in the
FIRST $\SI{21}{\cm}$ radio survey$^{30}$, the Two
Micron All Sky Survey (2MASS)$^{31}$, the Catalina Sky
Survey (CSS)$^{7}$, and in the \herschel, \wise\ and
\spitzer\ infrared satellite
archives$^{32,33,34}$.
Useful upper limits come from
non-detections in the Australia Telescope $\SI{20}{\giga\Hz}$ (AT20G)
survey$^{35}$ and the WISH
survey$^{36}$. Flux measurements, ranges (when time
resolved data are available) and upper limits from these sources are listed in
Extended Data Table~\ref{t:archive}.

We supplemented these data with our own intensive observations on a variety of
telescopes and instruments, namely: the $\SI{8.2}{\m}$ Very Large Telescope
(VLT) with the FORS and X-SHOOTER optical/NIR spectrographs and the HAWK-I NIR
imager; the $\SI{4.2}{\m}$ William Herschel Telescope (WHT) with the ISIS
spectrograph and the ULTRACAM high-speed camera$^{6}$;
the $\SI{2.5}{\m}$ Isaac Newton Telescope (INT) with the Intermediate
Dispersion Spectrograph (IDS); the $\SI{2.4}{\m}$ Thai National Telescope with
the ULTRASPEC high-speed camera$^{37}$;
the UV/optical and X-ray instruments UVOT and XRT on the \swift\
satellite; the COS UV spectrograph on the Hubble Space Telescope, \hst;
radio observations on the Australia Telescope Compact Array (ATCA).
Optical monitoring data came from a number of small telescopes. We include
here data taken with a $\SI{406}{\mm}$ telescope at the Remote
Observatory Atacama Desert (ROAD) in San Pedro de 
Atacama$^{38}$. Extended Data
Table~\ref{t:obslog} summarises these observations.

\subsection{The orbital, spin and beat frequencies.}
The orbital, spin and beat frequencies were best measured from the
small-telescope data because of their large time-base. For example, see the
amplitude spectrum around the spin/beat components of the clear filter data
from 19--28 July 2015 shown in Extended Data Fig.~\ref{f:ClearAmps}.  The
final frequencies, which give the dashed lines of Extended Data
Fig.~\ref{f:ClearAmps}, were obtained from the CSS data. These consisted of
305 exposures each 30-seconds in duration spanning the interval 30 May 2006
until 8 July 2013. We rejected 6 points which lay more than $4\sigma$ from
the multi-sinusoid fits that we now describe. To search for signals in these
sparsely-sampled data, we first transformed the UTC times of the CSS data to a
uniform timescale (TDB) and then corrected these for light-travel delays to
the solar system barycentre. The periodogram of these data is dominated by the
strong orbital modulation, which leaks so much power across the spectrum owing
to the sparse sampling that the spin/beat component can only be seen after the
orbital signal is removed. Once this was done, beat and spin components
matching those of Extended Data Fig.~\ref{f:ClearAmps} could be identified
(Extended Data Fig.~\ref{f:CRTSamps}). We carried out bootstrap multi-sinusoid
fits to compute the distributions of the orbital, beat and spin
frequencies. The orbital frequency closely follows a Gaussian distribution;
the beat and spin distributions are somewhat non-Gaussian in their high and low
frequency wings respectively, but are nevertheless well-defined. Statistics
computed from these distributions are listed in Extended Data
Table~\ref{t:freqs}.

Having established that the two pulsation frequencies are separated by the
orbital frequency, we carried out a final set of fits in which we enforced the
relation $\fspin-\fbeat=\forb$, but also allowed for a linear drift of the
pulsation frequency in order to be sensitive to any change in the pulsation
frequency. This led to a significant improvement in $\chi^2$ ($> 99.99$\%
significance on an $F$-test) which dropped from $326$ to $289$ for the 299
fitted points relative to a model in which the frequencies did not vary (after
scaling uncertainties to yield $\chi^2/N \approx 1$ for the final fit).
Bootstrap fits gave a near-Gaussian distribution for the frequency derivative
with $\dot{\nu}=-\SI{2.86(36)e-17}{\hertz\per\s}$.

Pulsations are detected at all wavelengths with suitable data other than
X-rays, where limited signal ($\approx 630$ source photons in
$\SI{10.2}{\kilo\s}$) leads to the upper limit of a $\SI{30}{\percent}$ pulse
fraction quoted in the main text. The \swift\ X-ray observations were taken in
$\SI{1000}{\s}$ chunks over the course of more than one month and we searched
for the pulsations by folding into 20 bins and fitting a sinusoid to the
result. There were no significant signals on either the beat or spin periods or
their harmonics. We used a power-law fit to the X-ray spectrum to deduce the
slopes shown in Fig.~\ref{f:sed}.

\subsection{The M star's spectral type and distance.}
The CSS data establish the orbital period $P =
\SI[separate-uncertainty=false]{0.14853528(8)}{\day}$, but not the absolute
phase of the binary. This we derived from observations of the M star, which
also led to a useful constraint upon the distance to the system.  The VLT+FORS
data were taken shortly before the photometric minimum, allowing a clear view
of the M star's contribution. We used M star spectral-type templates developed
from SDSS spectra$^{39}$ to fit AR~Sco's spectrum,
applying a flux scaling factor $\alpha$ to the selected template and adding a
smooth continuum to represent any extra flux in addition to the M star. The
smooth spectrum was parameterised by $\exp(a_1+a_2\lambda)$ to ensure
positivity. The coefficients $a_1$, $a_2$ and $\alpha$ were optimised for each
template, with emission lines masked since they are not modelled by the smooth
spectrum. Out of the templates available (M0-9 in unit steps), the M5 spectrum
gave by far the best match with $\chi^2=24$,$029$ for 1165 points fitted
compared to $>100$,$000$ for the M4 and M6 templates on either side (Extended
Data Fig.~\ref{f:spectype}).  The templates used were normalised such that the
scaling factor $\alpha=(R_2/d)^2$. We found $\alpha=3.02\times10^{-21}$, so
$R_2/d=5.5\times10^{-11}$. Assuming that the M star is close to its Roche lobe
(there is evidence supporting this assumption in the form of ellipsoidal
modulations of the minima between pulsations in the HAWK-I data,
Fig.~\ref{f:lc}), its mean density is fixed by the orbital period, which means
that its radius is fixed by its mass. Assuming $M_2=\SI{0.3}{\msun}$, for
reasons outlined in the main text, we find that $R_2=\SI{0.36}{\rsun}$, and
hence $d=\SI{149}{\parsec}$. This is an overestimate as the FORS spectrum was
taken through a narrow slit. We estimated a correction factor by calculating
the $i'$-band flux of the spectrum ($\SI{2.50}{\mjy}$) and comparing it to the
mean $i'$-band flux ($\SI{4.11}{\mjy}$) of the ULTRACAM photometry over the
same range of orbital phase. This is approximate given that the ULTRACAM data
were not taken simultaneously with the FORS data and there may be stochastic
variations in brightness from orbit-to-orbit, however the implied
$\SI{61}{\percent}$ throughput is plausible given the slit width of $0.7"$ and
seeing of $\sim 1"$. The final result is the distance quoted in the main text
of $d=\SI{116\pm16}{\parsec}$, and allows for uncertainties in the calibration
of the surface brightness of the templates and in the slit-loss correction.

We used the radius, spectral type and distance to estimate the $K_S$ flux
density from the donor as $f_{Ks}=\SI{9.4}{\mjy}$. The minimum observed flux
density from the HAWK-I data is $\SI{9.1}{\mjy}$. Uncertainties in the
extrapolation required to estimate the $K_S$ flux and from ellipsoidal
modulations allow the numbers to be compatible, but they suggest that the
estimated distance is as low as it can be and that the M star dominates the
$K_S$ flux at minimum light. The estimated M star fluxes for $i'$ and $g'$,
$f_{i'} = \SI{1.79}{\mjy}$ and $f_{g'}=\SI{0.07}{\mjy}$, are comfortably less
than the minimum observed fluxes of $\SI{2.57}{\mjy}$ and $\SI{0.624}{\mjy}$
in the same bands. We do not detect the white dwarf. The strongest constraint
comes from the \hst\ far ultraviolet data which at its lowest require
$T_1<\SI{9750}{\K}$. A white dwarf model atmosphere of $T=\SI{9750}{\K}$,
$\log g = 8$, corrected for slit-losses is plotted in Fig.~\ref{f:spectype},
and also (without slit losses) in Fig.~\ref{f:hstspec} which shows the average
\hst\ spectrum. Given the small maximum contribution of the white dwarf seen
in these figures, the absence of absorption features from the white dwarf is
unsurprising.

\subsection{The M star's radial velocity.}
We used spectra taken with the ISIS
spectrograph on the WHT and X-SHOOTER on the VLT to measure
radial velocities of the M star using the NaI 8200 doublet lines.
These vary sinusoidally on the same
$\SI{3.56}{\hour}$ period as the slowest photometric variation (Fig.~\ref{f:rvs}), hence our identification of this period as the orbital
period. We fitted the velocities with
\[V_R=\gamma+K_2\sin\left(2\pi(t-T_0)/P\right),\]
with the period fixed at the value obtained from the CSS data,
$P=\SI{0.14853528}{\day}$, and the systemic offset
$\gamma$ allowed to float free for each distinct subset of the data to allow
for variable offsets. We found $K_2=\SI{295\pm4}{\km\per\s}$ and
$T_0=\SI[separate-uncertainty=false]{57264.09615(33)}{\day}$, thus the orbital
ephemeris of AR~Sco is
\[\mathrm{BMJD(TDB)}=57264.09615(33)+0.14853528(8)E,\]
where $E$ is the
cycle number, and the time scale is TDB, corrected to the barycentre of the
solar system, expressed as a Modified Julian Day number ($\mathrm{MJD} =
\mathrm{JD} -2400000.5$). This ephemeris is important in establishing the origin of the
emission lines, as will be shown below.

The radial velocity amplitude and orbital period along with Kepler's third law
define the ``mass function''
\[\frac{M_1^3 \sin^3i}{(M_1+M_2)^2}=\frac{PK_2^3}{2\pi G}=\SI{0.395\pm0.016}{\msun},\]
where $i$ is the orbital inclination. This is a hard lower limit to the mass
of the compact object, $M_1$, which is only met for $i = 90^\circ$ and
$M_2=0$. There is however a caveat to this statement: it is sometimes observed
that irradiation can weaken the absorption lines on the side of the cool star
facing the compact object causing the observed radial velocity amplitude to be
an over-estimate of the true
amplitude$^{40,41}$. If this effect
applied here, which we suspect it might, both $K$ and the mass function limit
would need to be reduced. Given the large intrinsic variability of AR~Sco, and
the lack of flux-calibrated spectra, it was not possible to measure the
absolute strength of NaI. We attempted therefore to search for the influence
of irradiation from another side effect, which is that it causes the radial
velocity to vary non-sinusoidally$^{42}$. We failed to
detect any obvious influence of irradiation through this method, but its
effectiveness may be limited by the heterogeneous nature of our data which
required multiple systematic velocity offsets. Despite our failure to detect
clear signs of the effect of irradiation upon the M star's radial velocities,
we would not be surprised if the true value of $K$ was anything up to $\sim
\SI{20}{\km\per\s}$ lower than we measure. However, with no clear evidence for
the effect, in this we paper we proceed on the basis that we have measured the
true value of the M star's centre of mass radial velocity amplitude. This is
conservative in the sense that any reduction in $K$ would move the mass limits
we deduce to lower values, which would tilt the balance even more heavily
towards a white dwarf as the compact star. The UV and optical emission lines
come from the irradiated face of the M star and their amplitude compared to
$K_2$ sets a lower limit to the relative size of the M star, and hence,
through Roche geometry, the mass ratio $q = M_2/M_1$. Extended Data
Fig.~\ref{f:roche} shows how the emission measurements lead to the quoted
limit of $q > 0.35$, which leads in turn to the lower limits
$M_1>\SI{0.81}{\msun}$ and $M_2>\SI{0.28}{\msun}$ quoted in the main text.

The orbital period of a binary star sets a lower limit on the mean densities
of its component stars$^{43}$. Since the mean densities
of main-sequence stars decrease with increasing mass, this implies that we can
set an upper limit to the mass of any main-sequence component. In the case of
AR~Sco we find that $\left<\rho_2\right>>\SI{8900}{\kg\per\m\cubed}$ which leads
to $M_2<\SI{0.42}{\msun}$; the slightly larger value of $\SI{0.45}{\msun}$
quoted in the text allows for uncertainty in the models. The limit becomes an
equality when the M star fills its Roche lobe, which we believe to be the
case, or very nearly so, for AR~Sco. However, we expect that even in this case
the number deduced still functions as an upper limit because the mass-losing
stars in close binaries are generally over-sized and therefore less dense than
main-sequence stars of the same mass$^{44}$. Indeed,
systems with similar orbital periods to that of AR~Sco have donor star
masses in the range $\SI{0.2}{\msun}$ to
$\SI{0.3}{\msun}$$^{44}$. This, and the M5 spectral
type, are why we favour a mass of $M_2 \approx\SI{0.3}{\msun}$, close to the
lower limit on $M_2$.

\subsection{Brightness temperature at radio wavelengths.}
The pulsations in radio flux are a remarkable feature of AR~Sco, unique
amongst known white dwarfs and white dwarf binaries. If we assume that, as
at other wavelengths, and as suggested by the alignment of the second harmonic
power with $2\fbeat$ (Extended Data Fig.~\ref{f:pgram}), they arise largely
from the M star, then we can deduce brightness temperatures from the relation
\[ T_b = \frac{\lambda^2}{2\pi k} \left(\frac{d}{R_2}\right)^2 f_\nu .\] These
work out to be $\approx \SI{e12}{\K}$ and $\approx \SI{e13}{\K}$ for the
observations at $\nu = \SI{5.5}{\GHz}$ and $\nu = \SI{1.4}{\GHz}$
respectively. Although the value at the lowest frequency exceeds the $\sim
\SI{e12}{\K}$ limit at which severe cooling of the electrons due to inverse
Compton scattering is thought to occur$^{45}$, this is
not necessarily a serious issue given the short-term variability exhibited by
the source. The limits can be lowered by appealing to a larger emission region
as the radio data in hand are not enough to be certain that emission arises
solely on the M star. Even so, the $\SI{0.98}{\minute}$ second harmonic
pulsations that are seen in the radio flux suggest an upper limit to the size
of the emission region of $\SI{25}{\rsun}$ from light-travel time alone. This
implies a minimum brightness temperature of $\SI{e9}{\K}$ at $\SI{1.4}{\GHz}$,
showing clearly that the radio emission is non-thermal in origin. We assume
that synchrotron emission dominates; while there may be thermal and cyclotron
components at shorter wavelengths, there is no clear evidence for either.

\subsection{Code availability.}
The data were reduced with standard instruments pipelines for the \hst, VLT,
and \swift\ data. The WHT and INT data were reduced with STARLINK software.
Scripts for creating the figures are available from the first author apart
from the code for computing the white dwarf model atmosphere, which is a
legacy F77 code and complex to export. The atmosphere model itself
however is available on request.

\end{methods}

\noindent
\begin{enumerate}
\setcounter{enumi}{29}
\item Becker, R.~H., White, R.~L. \& Helfand, D.~J. The FIRST Survey: Faint Images of the Radio Sky at Twenty Centimeters. \textit{Astrophys.\ J.\ } \textbf{450,} 559--577 (1995).

\item Skrutskie, M.~F. \textit{et al.} The Two Micron All Sky Survey (2MASS). \textit{Astron.\ J.\ } \textbf{131,} 1163--1183 (2006).

\item Pilbratt, G.~L. \textit{et al.} Herschel Space Observatory. An ESA facility for far-infrared and submillimetre astronomy. \textit{Astron.\ \& Astrophys.\ } \textbf{518,} L1--L6 (2010).

\item Wright, E.~L. \textit{et al.} The Wide-field Infrared Survey Explorer (WISE): Mission Description and Initial On-orbit Performance. \textit{Astron.\ J.\ } \textbf{140,} 1868--1881 (2010).

\item Werner, M.~W. \textit{et al.} The Spitzer Space Telescope Mission. \textit{Astrophys.\ J.\ Supp.\ } \textbf{154,} 1--9 (2004).

\item Murphy, T. \textit{et al.} The Australia Telescope 20 GHz Survey: the source catalogue. \textit{Mon.\ Not.\ R.\ Astron.\ Soc.\ } \textbf{402,} 2403--2423 (2010).

\item De Breuck, C., De Breuck, C., De Breuck, C., R\"ottgering, H. \& van Breugel, W. A sample of ultra steep spectrum sources selected from the Westerbork In the Southern Hemisphere (WISH) survey. \textit{Astron.\ \& Astrophys.\ } \textbf{394,} 59--69 (2002).

\item Dhillon, V.~S. \textit{et al.} ULTRASPEC: a high-speed imaging photometer on the 2.4-m Thai National Telescope. \textit{Mon.\ Not.\ R.\ Astron.\ Soc.\ } \textbf{444,} 4009--4021 (2014).

\item Hambsch, F.-J. ROAD (Remote Observatory Atacama Desert): Intensive Observations of Variable Stars. \textit{Journal of the American Association of Variable Star Observers (JAAVSO)} \textbf{40,} 1003--1009 (2012).

\item Rebassa-Mansergas, A., Rebassa-Mansergas, A., Rebassa-Mansergas, A., Schreiber, M.~R. \& Koester, D. Post-common-envelope binaries from SDSS - I. 101 white dwarf main-sequence binaries with multiple Sloan Digital Sky Survey spectroscopy. \textit{Mon.\ Not.\ R.\ Astron.\ Soc.\ } \textbf{382,} 1377--1393 (2007).

\item Hessman, F.~V., Hessman, F.~V., Nather, R.~E. \& Zhang, E.-H. Time-resolved spectroscopy of SS Cygni at minimum and maximum light. \textit{Astrophys.\ J.\ } \textbf{286,} 747--759 (1984).

\item Wade, R.~A. \& Horne, K. The radial velocity curve and peculiar TiO distribution of the red secondary star in Z Chamaeleontis. \textit{Astrophys.\ J.\ } \textbf{324,} 411--430 (1988).

\item Marsh, T.~R. A spectroscopic study of the deeply eclipsing dwarf nova IP Peg. \textit{Mon.\ Not.\ R.\ Astron.\ Soc.\ } \textbf{231,} 1117--1138 (1988).

\item Faulkner, J., Flannery, B.~P. \& Warner, B. Ultrashort-Period Binaries. II. HZ 29 (=AM CVn): a Double-White Semidetached Postcataclysmic Nova?. \textit{Astrophys.\ J.\ Lett.\ } \textbf{175,} L79--L83 (1972).

\item Knigge, C., Baraffe, I. \& Patterson, J. The Evolution of Cataclysmic Variables as Revealed by Their Donor Stars. \textit{Astrophys.\ J.\ Supp.\ } \textbf{194,} 28--75 (2011).

\item Kellermann, K.~I. \& Pauliny-Toth, I.~I.~K. The Spectra of Opaque Radio Sources. \textit{Astrophys.\ J.\ Lett.\ } \textbf{155,} L71--L78 (1969).

\end{enumerate}

\end{refsegment}


\begin{efigure}
  \begin{center}
    \includegraphics[width=0.9\textwidth]{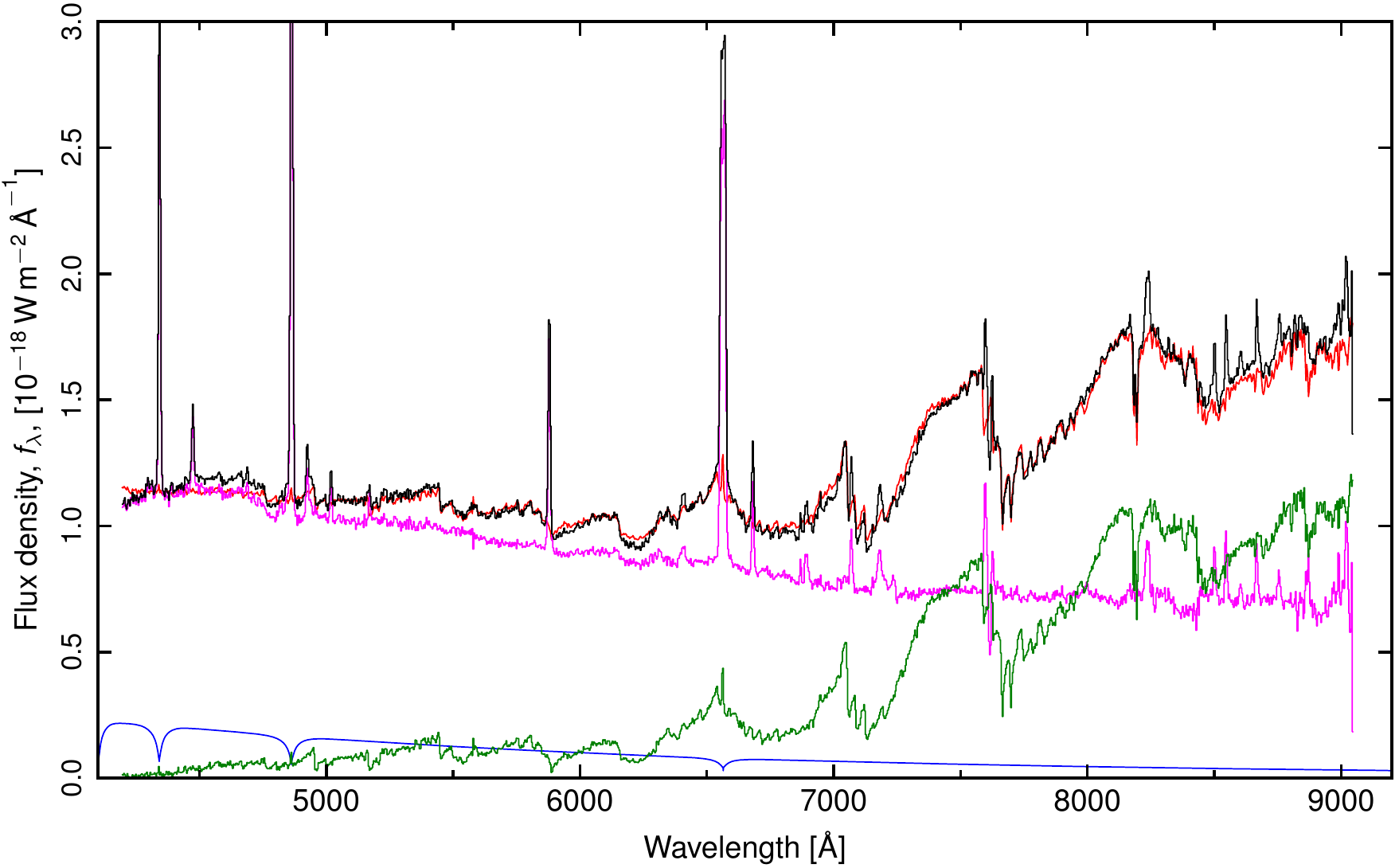}
  \end{center}
  \caption{\textbf{The optical spectrum of the white dwarf's M star
      companion.}  A 10~minute exposure of AR~Sco taken with FORS on the VLT
    between orbital phases 0.848 and 0.895 (black). Other spectra: an
    optimally-scaled M5 template (green); the sum of the template plus a
    fitted smooth spectrum (red); AR~Sco minus the template, i.e. the extra
    light (magenta); a white dwarf model atmosphere of $T = \SI{9750}{\K}$,
    $\log g = 8.0$, the maximum possible consistent with the \hst\ data
    (blue). A slit-loss factor of $0.61$ has been applied to the models. The
    strong emission lines come from the irradiated face of the M star.
    \label{f:spectype}}
\end{efigure}

\begin{efigure}
  \begin{center}
    \includegraphics[width=0.9\textwidth]{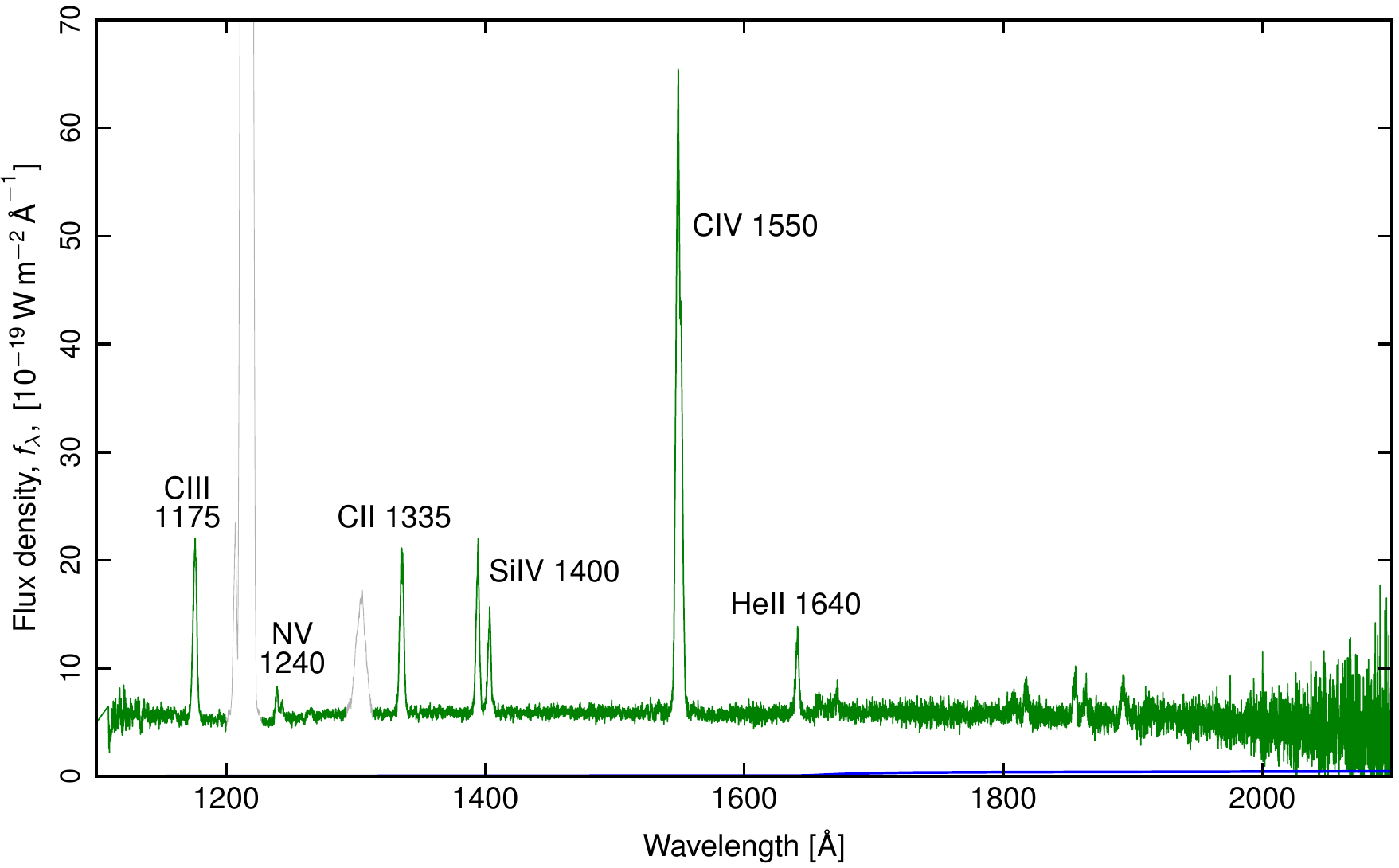}
  \end{center}
  \caption{\textbf{\hst\ ultraviolet spectrum of AR~Sco.} This shows the mean
    \hst\ spectrum with geocoronal emission plotted in grey. The blue line
    close to the $x$-axis is a white dwarf model atmosphere of $T =
    \SI{9750}{\K}$, $\log g = 8.0$, representing the maximal contribution of
    the white dwarf consistent with light-curves. The radial velocities of the
    emission lines (Extended Data Fig.~\protect\ref{f:roche}) show that, like
    the optical lines, the ultraviolet lines mainly come from the irradiated
    face of the M star.\label{f:hstspec}}
\end{efigure}

\begin{efigure}
\begin{center}
\includegraphics[width=0.99\textwidth]{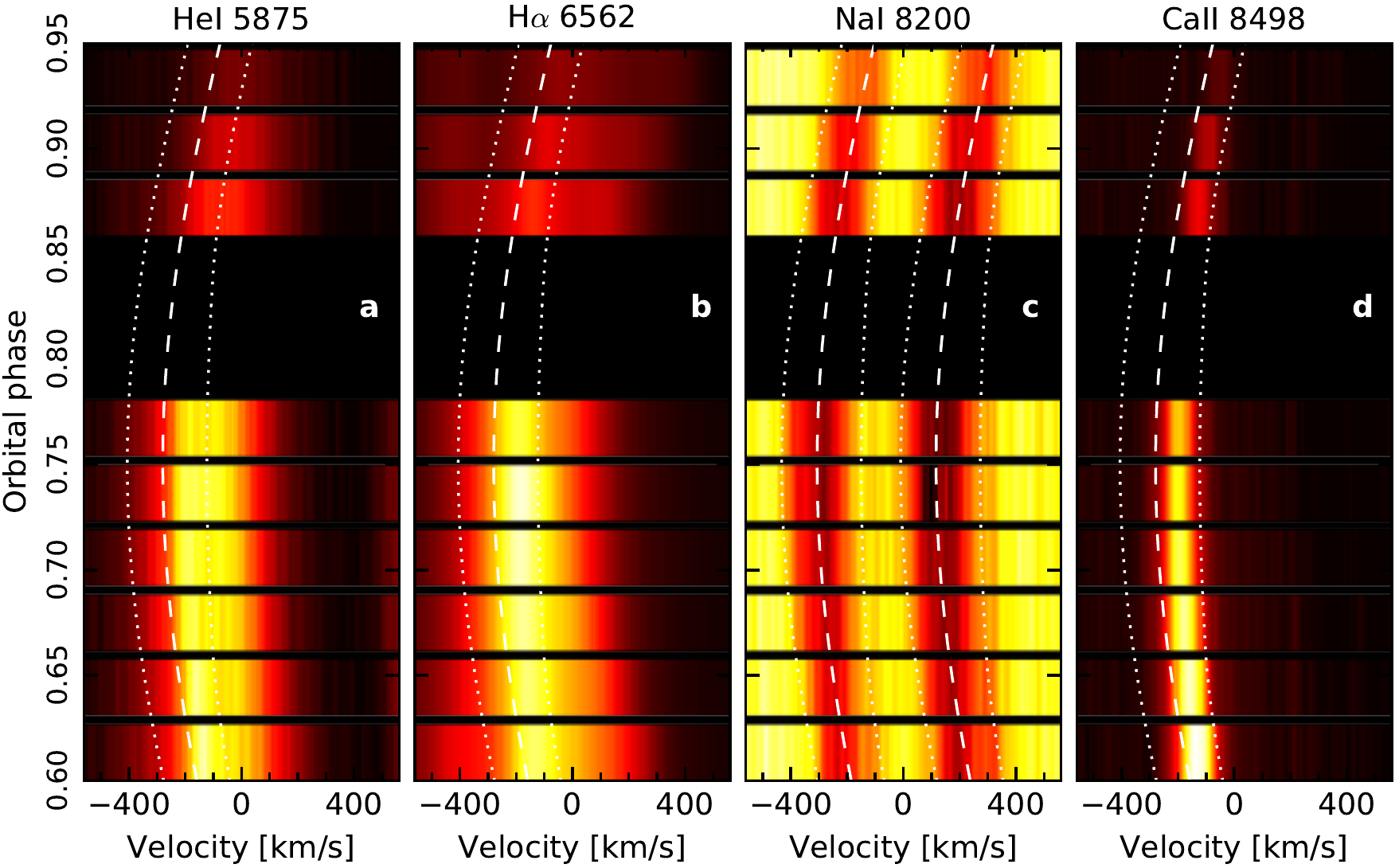}
\end{center}
\caption{\textbf{Velocity variations of atomic emission lines compared to
    those of the M star.} \textbf{a}, \textbf{b} and \textbf{d} show emission
  lines from a sequence of spectra from the VLT+X-SHOOTER data; \textbf{c}
  shows the NaI~8200 absorption doublet from the M star.  The dashed line
  shows the motion of the centre of mass of the M star deduced from the NaI
  measurements while the dotted lines show the maximum range of radial
  velocities from the M star for $q = M_2/M_1 =0.35$. The emission lines move
  in phase with the NaI doublet but at lower amplitude, showing that they come
  from the inner face of the M star.\label{f:trail}}
\end{efigure}

\begin{efigure}
  \begin{center}
    \includegraphics[width=0.8\textwidth]{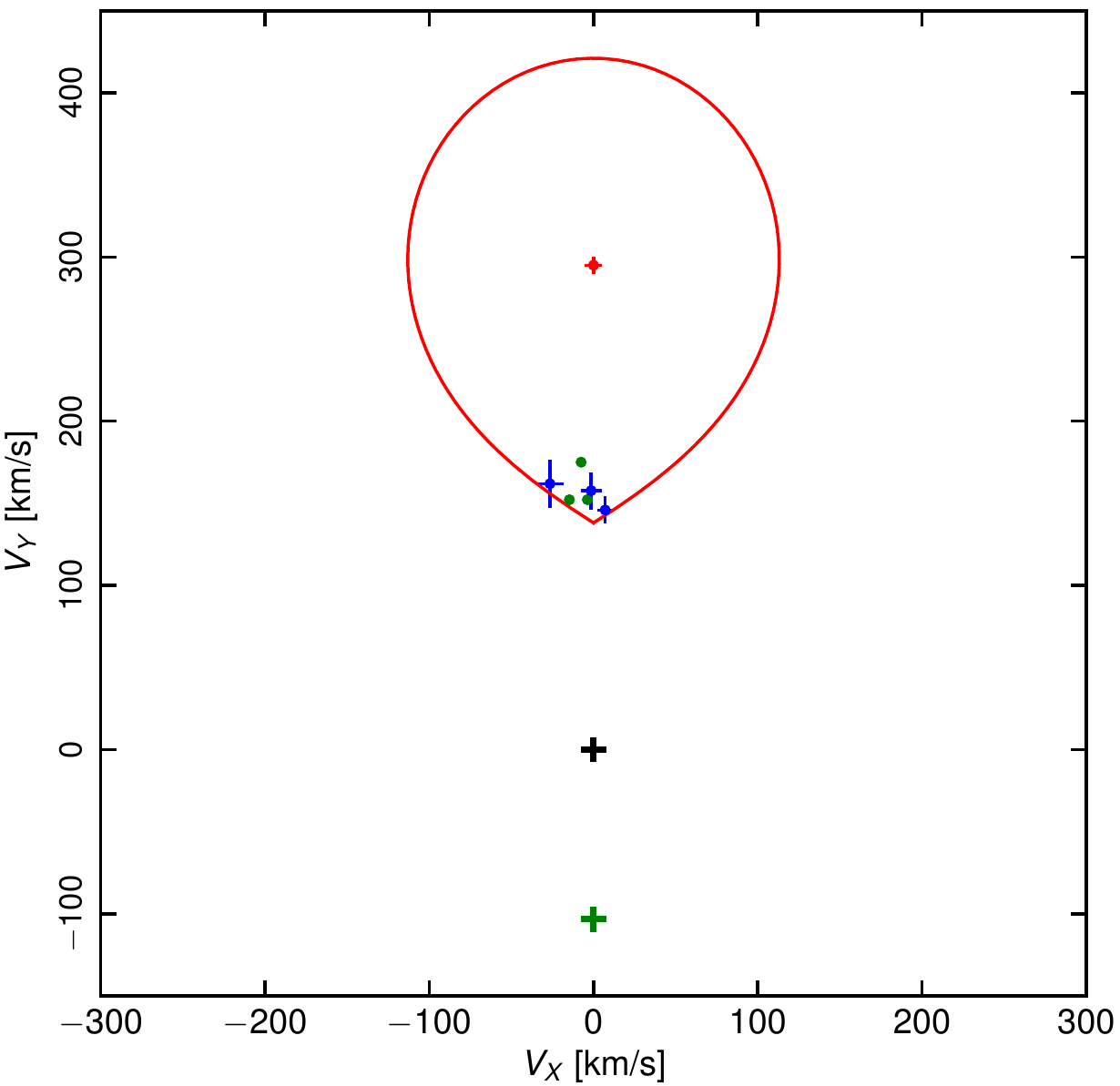}
  \end{center}
  \caption{\textbf{The emission lines' origin relative to the M star.}
    Velocities of the lines were fitted with $V_R = -V_X \cos 2\pi \phi + V_Y
    \sin 2\pi \phi$. The points show the values of $(V_X,V_Y)$. Red: the M
    star from NaI (by definition this lies at $V_X=0$). Blue: SiIV and HeII
    lines from the \hst\ FUV data. Green: H$\alpha$, $\beta$ and $\gamma$ from
    optical spectroscopy. The black and green plus signs mark the centres of
    mass of the binary and white dwarf respectively. The red line shows the
    Roche lobe of the M star for a mass ratio $q = 0.35$.\label{f:roche}}
\end{efigure}

\begin{efigure}
  \begin{center}
    \includegraphics[width=0.9\textwidth]{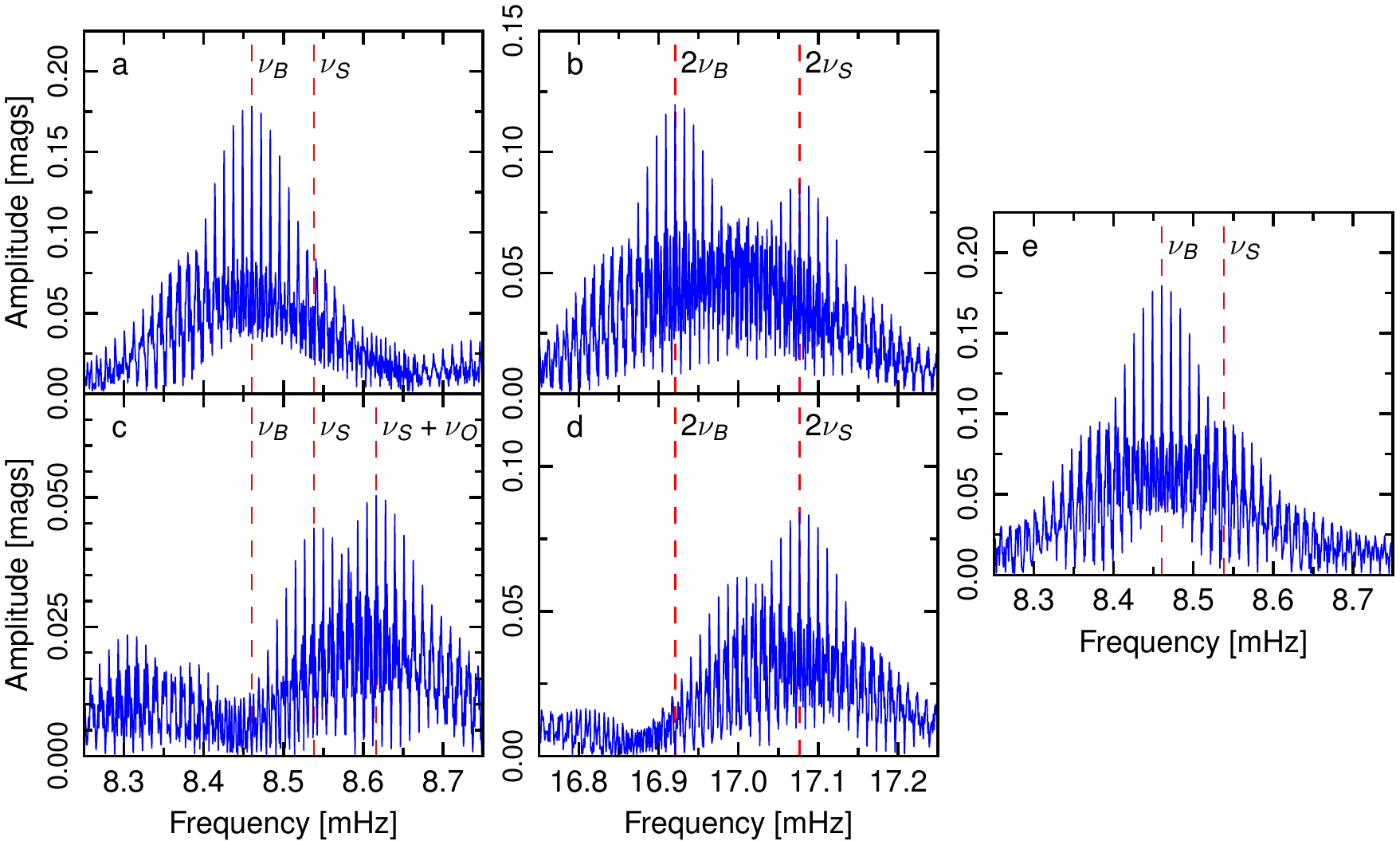}
  \end{center}
  \caption{\textbf{Amplitude spectra from 9 days monitoring with a small
      telescope.} \textbf{a}, Amplitude as a function of frequency around the
    $\SI{1.97}{\minute}$ signal from data taken with a $\SI{40}{\cm}$
    telescope. \textbf{b}, The same at the second harmonic. \textbf{c} and
    \textbf{d}, The same as \textbf{a} and \textbf{b} after subtracting the
    beat frequency signals at $\fbeat$ and $2\fbeat$. Signals at $\fspin +
    \forb$ and $2\fspin - \forb$ are also apparent. \textbf{e}, The window
    function, computed from a pure sinusoid of frequency $\fbeat$ and
    amplitude $0.18$ magnitudes (cf \textbf{a}).\label{f:ClearAmps}}
\end{efigure}

\begin{efigure}
  \begin{center}
    \includegraphics[width=0.9\textwidth]{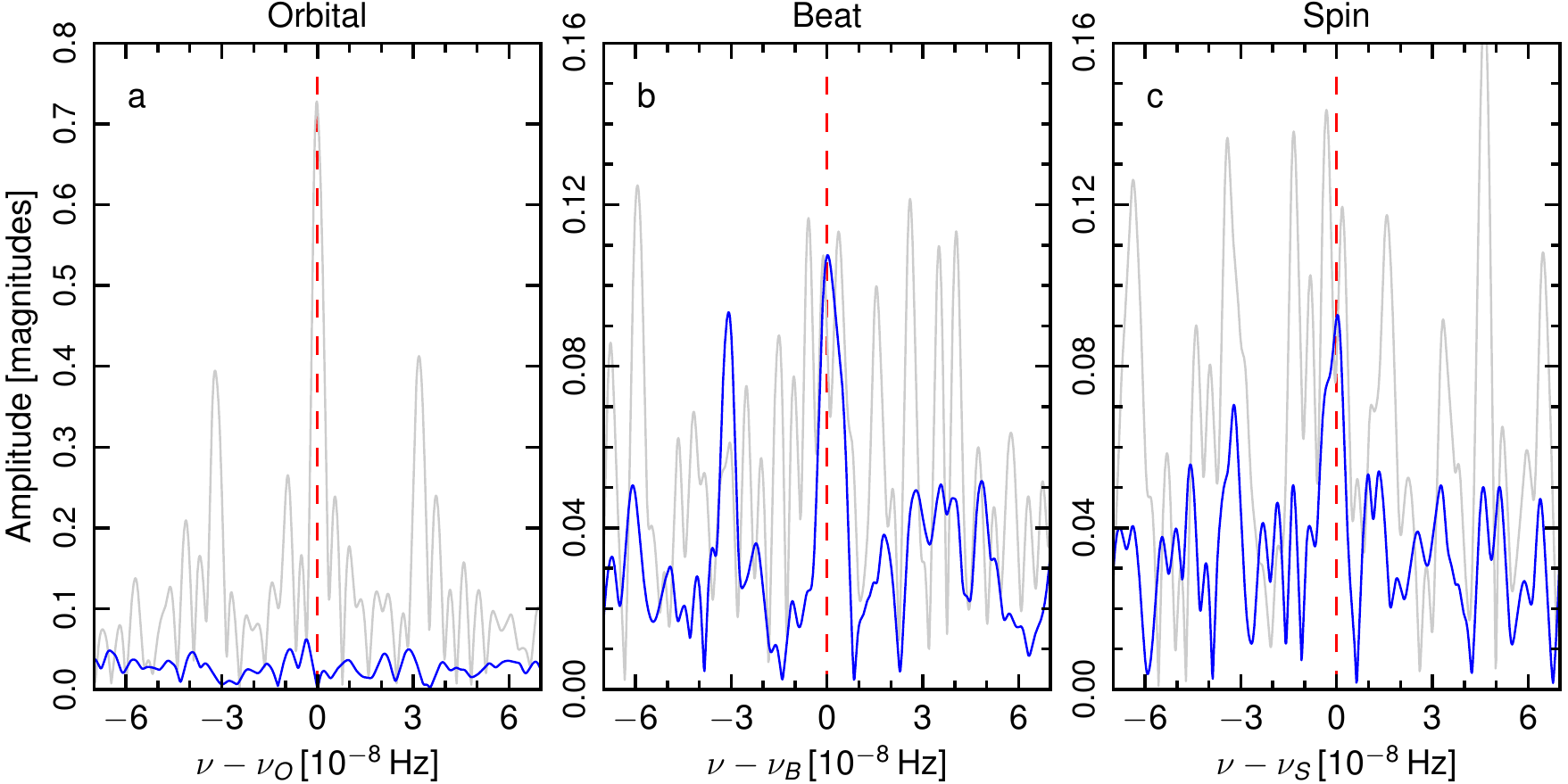}
  \end{center}
  \caption{\textbf{Amplitude spectra from 7 years of sparsely-sampled CSS
      data.} \textbf{a}-\textbf{c}, The amplitude as
    a function of frequency relative to the mean orbital (\textbf{a}), 
    beat (\textbf{b}) and spin (\textbf{c})
    frequencies listed in Extended Data Table~\protect\ref{t:freqs}.
    The grey line is the spectrum without any processing; the
    blue line is the spectrum after subtraction of the orbital signal.
    \label{f:CRTSamps}}
\end{efigure}


\newpage

\begin{table}
  \begin{center}
\begin{tabular}{lllp{1in}l}
\hline
Tel./Inst.  & Type                & Wavelength             & Date       & Exposure  \\
            &                     &                        &            &
            $T[\mathrm{s}]\times N$ \\
\hline
VLT+FORS    & Spectra    & $420$ -- $\SI{900}{\nm}$ & 2015-06-03 & 600x1    \\
WHT+ULTRACAM& Photometry & $u'$, $g'$, $r'$         & 2015-06-23 & 2.9x768    \\
WHT+ULTRACAM& Photometry & $u'$, $g'$, $i'$         & 2015-06-24 & 1.3x7634   \\
\swift+UVOT/XRT    & UV, X-rays& $\SI{260}{\nm}$, $0.2$ -- $\SI{10}{\keV}$ &
2015-06-23 -- 2015-08-03 & 1000x10\\
VLT+HAWKI   & Photometry & $K_S$                    & 2015-07-06 & 2.0x7020   \\
WHT+ISIS    & Spectra    & $354$ -- $539$, $617$ -- $\SI{884}{\nm}$ & 2015-07-16 & 20x94\\
WHT+ISIS    & Spectra    & $354$ -- $539$, $617$ -- $\SI{884}{\nm}$ & 2015-07-17 & 300x4\\
WHT+ISIS    & Spectra    & $356$ -- $520$, $540$ -- $\SI{697}{\nm}$ & 2015-07-19 & 30x130\\
ROAD $\SI{40}{\cm}$ & Photometry & White light & 2015-07-19 -- 2015-07-28 & 30x1932\\
WHT+ISIS    & Spectra    & $356$ -- $520$, $540$ -- $\SI{697}{\nm}$ & 2015-07-20 & 30x210\\
INT+IDS     & Spectra    & $440$ -- $\SI{685}{\nm}$                   & 2015-07-22 & 27x300\\
INT+IDS     & Spectra    & $440$ -- $\SI{685}{\nm}$                   & 2015-07-23 & 34x300\\
ATCA        & Radio      & $5.5$, $\SI{9.0}{\GHz}$                  & 2015-08-13 & 271x10\\
WHT+ISIS    & Spectra    & $320$ -- $535$, $738$ -- $\SI{906}{\nm}$ & 2015-08-26 & 600x8\\
WHT+ISIS    & Spectra    & $320$ -- $535$, $738$ -- $\SI{906}{\nm}$ & 2015-09-01 & 600x8\\
VLT+XSHOOTER& Spectra    & $302$ -- $\SI{2479}{\nm}$  & 2015-09-23 & 11x300\\
\emph{HST}+COS & Spectra & $110$ -- $\SI{220}{\nm}$   & 2016-01-19 & 5 orbits\\
TNT+ULTRASPEC & Photometry & $g'$ & 2016-01-19 & 3.8x1061\\

\hline
\end{tabular}
  \end{center}
\caption{\textbf{Observation log.} \label{t:obslog}}
\end{table}

\begin{table}
  \begin{center}
\begin{tabular}{lccccc}
\hline
Frequency         & $\SI{5}{\percent}$-ile & $\SI{95}{\percent}$-ile & Median & Mean & RMS\\
                  & mHz & mHz & mHz & mHz & mHz \\
\hline
$\forb$  & 0.077921311 & 0.077921449 & 0.077921380 & 0.077921380 &
0.000000042\\
$\fbeat$ & 8.4603102 & 8.4603140 & 8.4603112 & 8.4603114 & 0.0000011\\
$\fspin$ & 8.5382332 & 8.5382356 & 8.5382348 & 8.5382346 & 0.0000008\\
\hline
\end{tabular}
  \end{center}
  \caption{\textbf{Statistics of the orbital, beat and spin frequencies
      from bootstrap fits.} \label{t:freqs}}
\end{table}

\begin{table}
  \begin{center}
\begin{tabular}{llllll}
\hline
Source     & Wavelength,        & Flux  &
Source     & Wavelength,        & Flux  \\
           & Frequency          & mJy &
           & Frequency          & mJy \\
\hline
WISH       & $\SI{352}{\mega\Hz}$  & $< 18$ &
\wise\     & $\SI{22.0}{\micro\m}$  & $45.2$ -- $105.4$ \\
FIRST      & $\SI{1.4}{\giga\Hz}$  & $8.0\pm0.3$ &
\wise\     & $\SI{12}{\micro\m}$   & $18.0$ -- $48.3$ \\
AT20G      & $\SI{20}{\giga\Hz}$   & $< 50$ &
\spitzer\  & $\SI{5.73}{\micro\m}$ & $11.9$ -- $23.5$ \\
\herschel\ & $\SI{500}{\micro\m}$  & $92 \pm 25$ &
\wise\     & $\SI{4.60}{\micro\m}$ & $11.8$ -- $20.5$ \\
\herschel\ & $\SI{350}{\micro\m}$  & $76 \pm 21$ &
\spitzer\  & $\SI{3.6}{\micro\m}$  & $13.0 \pm 0.7$ \\
\herschel\ & $\SI{250}{\micro\m}$  & $55 \pm 23$ &
\wise\     & $\SI{3.4}{\micro\m}$  & $13.2$ -- $13.8$ \\
\herschel\ & $\SI{160}{\micro\m}$  & $118 \pm 38$ &
2MASS      & $\SI{2.1}{\micro\m}$  & $13.5 \pm 0.3$ \\
\herschel\ & $\SI{70}{\micro\m}$   & $196 \pm 63$ &
2MASS      & $\SI{1.7}{\micro\m}$  & $15.0 \pm 0.3$ \\
\spitzer\  & $\SI{24}{\micro\m}$   & $59.9 \pm 6.0$ &
2MASS      & $\SI{1.2}{\micro\m}$  & $13.3 \pm 0.3$ \\
\hline
\end{tabular}
  \end{center}
\caption{\textbf{Archival data sources and flux values.}\label{t:archive}}
\end{table}

\end{document}